\documentclass[a4paper]{jpconf}
\usepackage{graphicx,amssymb}
\begin{document}
\title{On the groundstate of octonionic matrix models in a ball}
\author{L Boulton$^{1}$, M P Garcia del Moral$^{2}$ and A Restuccia$^{2}$}

\address{$^1$ Maxwell Institute for Mathematical Sciences and Department of Mathematics, Heriot-Watt University, Edinburgh, EH14 4AS, United Kingdom.\\
$^2$ Departamento de F\'isica, Universidad de Antofagasta, Antofagasta, Aptdo 02800, Chile.}

\ead{L.Boulton@hw.ac.uk; maria.garciadelmoral@uantof.cl; alvaro.restuccia@uantof.cl}

\begin{abstract}
In this work we examine the existence and uniqueness of the groundstate of a $SU(N)\times G_2$ octonionic matrix model on a bounded domain of $\mathbb{R}^N$. The existence and uniqueness argument  of the groundstate wavefunction follows from the Lax-Milgram theorem. Uniqueness is shown by means of an explicit argument which is drafted in some detail.

\end{abstract}

\section{Introduction}
Matrix model groundstate wavefunctions have been investigated by means of different approaches in the past. In the quest for a better understanding of M-theory, this has been carried out either from the point of view of Supermembrane Theory \cite{dwhn},\cite{hoppe},\cite{fh}; or from the point of view of matrix models \cite{bfss}; or from the point of view of Yang-Mills theories described in the slow mode regime \cite{halpern}. The existence of a groundstate has been tested indirectly by means of a unique normalizable zero-energy wavefunction invariant under $SO(9)\times SU(N)$ \cite{yi},\cite{sethi-stern},\cite{porrati}. Direct attempts to characterize it include those reported in \cite{dwhn},\cite{hoppe}. Regimes near the origin were examined in \cite{hlt2} and asymptotic regimes were considerd in \cite{halpern} for different supersymmetric matrix models. The interest in a better understanding of theses theories relies in part on the AdS/CFT conjecture \cite{maldacena}.

The $N=16$ supersymmetric $SU(N)$ matrix model is dual to the decoupling limit of the D0-brane geometry in the type IIA String Theory \cite{ydri,connor}. The spectrum of this theory is continuous from zero to infinity as it was established in \cite{dwln}. In the BFSS interpretation, the nonzero energy eigenstates of the model correspond to scattering states that form a continuum. In the
bulk picture, black holes can decay into radiating D0-branes that can escape to infinity \cite{lin-yin}. The D0-branes are long-life metastable states associated in the bulk description to the
microstates of a black hole at finite temperatures.

A rigorous treatment of the existence and uniqueness of the groundstate of these matrix models would contribute towards a better understanding of the nature of these matrix model metastable states. Several authors have attempted to characterize the asymptotic  structure of
the groundstate wave function \cite{fghhy}. An asymptotic expansion for the groundstate wave
function in the $SU(2)$ case has been studied in \cite{su(2)}. Subsequent proposals for
$N= 3$ were made in \cite{hp}. For arbitrary $N$ a more recent analysis has been considered in \cite{lin-yin}.

In this note we examine the existence and uniqueness of a groundstate of a supersymmetric matrix model with $G2$ symmetry on a compact space. The latter corresponds to a $SO(7)$ truncation of the supermembrane analyzed in \cite{dwhn}. As it was shown in \cite{dwhn}, this model  does not admit a normalizable state when the spacetime is noncompact.

Our interest for examining this model on a bounded domain is two-folded. On the one hand, it provides a suitable illustration of new techniques for determining the existence of massless groundstates for a class of supersymmetric bounded matrix models. On the other hand, it has an intrinsic interest as an example of octonionic supersymmetric quantum mechanics \cite{dd}.

Octonion quantum mechanics was introduced in the context of strong interactions and decay \cite{oqm}. A new quantum theory described in terms of  nonassociative but commutative algebra was introduced as a basic tool in \cite{jordan}. Yang Mills theories valued on the octonionic algebra with gauge group G2, the automorphisms of the octonions, was recently analysed in \cite{veiro-restuccia}.

Despite of the fact that the decay description in Fermi theory and the Yukawa model of nuclear force rendered this type of construction useless in its original context, these models have remained of substantial interest in the theory of matrix models. The automorphisms of the octonions is the group $G2$, which is well-known to contain the Standard Model. A supermembrane with octonionic twists and a gauge $G2\times U(1)\times SU(N)$ symmetry
was considered in \cite{hlt}.
The symmetry in the latter was obtained by performing a deformation of the original $11D$ supermembrane matrix model described in a $11D$ Minkowski spacetime
theory \cite{dwhn}.

\section{ The truncated octonionic $D=11$ supermembrane}

This example of truncation of the supermembrane in the L.C.G. was originally formulated \cite{dwhn} in a Minkowski spacetime $M_9$. Although it is possible to obtain explicitly two solutions for the ground state problem, it was shown in \cite{dwhn} that they fail to be square-integrable so the truncated model has no massless states. In the following we examine the groundstate wavefunction, but now for the restricted case when the spatial part of the target space is  compact.

The model may be formulated in terms of a pure imaginary octonion with coefficients valued on the $su(N)$ algebra:
\[X=X^{A}_{i}e^iT_A,\]
where $X_i^A$ are real $(0+1)$ fields which only depend on the time coordinate, $T_A$ are the generators of the $su(N)$ algebra, $A=1,\dots,N^2-1$ and  $e_i$ denote the pure imaginary basis of the octonionic non-associative division algebra.

The other object involved in the supersymmetric model is a pure imaginary octonion with coefficients valued on the $su(N)$ algebra
\[
\lambda= \lambda^A_i e^iT_A
\]
where $\lambda^A_i$ are $(1+0)$ fields valued on an odd part of a Grassmanian algebra. \\

The quantum hamiltonian is given by
\[H= H_b+H_f. \]
The bosonic part of the hamiltonian is a Schr\"odinger operator
\[ H_b=-\frac{1}{2}\Delta+V(X)\]

where
\[V(X)=\frac{1}{4}f_{ABE}f_{CDF}\delta^{EF}X^A_iX^B_jX^C_iX^D_j,\]
$f_{ABE}$ are the structure constants of $su(N)$ algebra
\[\left[T_A,T_B\right]=f_{ABE}\delta^{EC}T_C.\]
The fermionic part of the hamiltonian is given by
\[H_f=-f_{ABD}X^A_i\lambda^B_j c^{ijk}\frac{\partial}{\partial\lambda^D_k}\]
where $c^{ijk}$ are the structure constants $[e^i,e^j]=c^{ijk}e^k$ of the octonionic algebra. The bosonic potential $V(X)$ is a quartic polynomial in $X$, while the fermionic hamiltonian is linear in the $X$ variable.

These are the characteristic properties of the hamiltonian of the $D=11$ supermembrane. The hamiltonian is invariant under the group $G_2$, the automorphisms of the octonions, and under $SU(N)$ associated to the regularized model. These are rigid symmetries. The hamiltonian is also invariant under $N=1$ supersymmetry with the generators:
\begin{equation}
Q=\left(\frac{\partial}{\partial X^A_i}+\frac{1}{2}c^{ijk}f_{ABC}X^B_jX^C_k\right)\lambda^A_i\qquad
Q^{\dag}=\left(-\frac{\partial}{\partial X^A_i}+\frac{1}{2}c^{ijk}f_{ABC}X^B_jX^C_k\right)\frac{\partial}{\partial\lambda^A_i}.
 \end{equation}
The corresponding anticommutation relations are:
\[\{Q,Q\}=\{Q^{\dag},Q^{\dag}\}=0,\quad \{Q,Q^{\dag}\}=2H.
\]

\section{The Dirichlet problem for the hamiltonian $H$ on a compact domain $\Omega$}

Let $\Omega$ be a bounded domain with piecewise smooth boundary. The problem can be formulated as follows: find a wavefunction $\Psi$ such that $H\Psi=0$ in $\Omega$, satisfying the boundary condition $\Psi=f$ on $\partial\Omega$.

Here $\Psi$ is a superfield, that is, it admits an expansion in terms of the Grassmann elements as follows:
\[\Psi=\Psi_0(X)+\Psi^i_{1A}\lambda^A_{i}
+\Psi^{ij}_{2A_1A_2}(X)\lambda^{A_1}_{i}\lambda^{A_2}_{j}+\dots
+\Psi^{i_1\dots i_m}_{mA_1,\dots A_m}(X)\lambda^{A_1}_{i_1}\dots\lambda^{A_m}_{i_m}+\dots
\]
This is a finite expansion, since $A=1,\dots,N^2-1$ and $i=1,\dots,7$,
and $\lambda$ are odd elements of the Grassman algebra.

\subsection{Existence of the solution}
Since the hamiltonian $H$ is a Schr\"odinger operator on a compact domain with a smooth potential, existence of the wavefunction $\Psi$ follows from the Lax-Milgram theorem.

\subsection{Uniqueness of the solution}
Uniqueness can be shown along the following lines. If $\Psi_1$ and $\Psi_2$ are two solutions, then $H(\Psi_1-\Psi_2)=0$ in $\Omega$ and $\Psi_1-\Psi_2=0$ on $\partial\Omega$. We then show that $H\Psi=0$ in $\Omega$ and $\Psi=0$ on $\partial\Omega$, imply $\Psi=0$ in $\Omega$. Hence $\Psi_1=\Psi_2$ in $\Omega$ will follow immediately.

For simplicity we assume that $\Omega$ is an hypercube with faces given by the hypersurface obtained by fixing a constant value for one of the coordinates, say $X^{\hat{A}}_{\hat{i}}=(\mathrm{ctt})$. We remark that the following argument may also be implemented on any $\Omega$ with smooth boundary.

Firstly observe that
\[
H\Psi=0 \qquad \mathrm{implies} \qquad \left\{  \begin{array}{c} Q\Psi=0 \\  Q^{\dag}\Psi=0 \end{array} \right. \qquad(\textrm{in} \quad \Omega).
\]
Below we employ this property of the supersymmetric hamiltonian.

Assume $Q\Psi=0$ in $\Omega$. Then, when we approach to a point on the hypersurface $X^{\hat{A}}_{\hat{i}}=(\mathrm{ctt})$, due to continuity and the boundary condition $\Psi=0$, we have
\[
\frac{\partial}{\partial X^{A}_i}\lambda^{A}_i\Psi=0.
\]
At the surface $X^{\hat{A}}_{\hat{i}}=(\mathrm{ctt})$, the partial derivatives with respect to the other coordinates vanish. Hence $\frac{\partial\Psi}{\partial X_i^A}=0$ for $(A,i)\neq(\hat{A},\hat{i})$. We are then left with
\[
\lambda^{\hat{A}}_{\hat{i}}\frac{\partial\Psi_{0}(X)}{\partial X^{\hat{A}}_{\hat{i}}}
+\lambda^{\hat{A}}_{\hat{i}}\frac{\partial\Psi^{j_1}_{1B_1}(X)}{\partial X^{\hat{A}}_{\hat{i}}}\lambda^{B_1}_{j_1}+\dots
+\lambda^{\hat{A}}_{\hat{i}}\frac{\partial\Psi^{j_1\dots j_m}_{mB_1\dots B_m}(X)}{\partial X^{\hat{A}}_{\hat{i}}}\lambda^{B_1\dots B_m}_{j_1\dots jm}=0
\]
Since each term has a different number of $\lambda$ factors, then each term must be zero. We then have from the first terms
\begin{equation}\label{I}\frac{\partial\Psi_{0}(X)}{\partial X^{\hat{A}}_{\hat{i}}}=0.
\end{equation}
\begin{equation}\label{II}\frac{\partial\Psi^{j}_{1B}(X)}{\partial X^{\hat{A}}_{\hat{i}}}
=0\quad \textrm{for }(B,j)\ne(\hat{A},\hat{i}).
\end{equation}
When $(B,j)=(\hat{A},\hat{i})$ there is no restriction on the corresponding wavefunction (from the anticommuting properties of $\lambda$).
For the generic term we obtain
\begin{equation}\label{III}\frac{\partial\Psi^{j_1\dots j_m}_{m B_1\dots B_m}(X)}{\partial X^{\hat{A}}_{\hat{i}}}=0
\end{equation}
for the indices $(B_1,j_1),(B_2,j_2),\dots,(B_m,j_m)$ different from $(\hat{A},\hat{i}).$
Without loss of generality we can assume  $\Psi^{j_1,\dots,j_m}_{m B_1,\dots,B_m}(X)$ to be antisymmetric under the exchange of the $(B,j)$ indices.

Now, assume that $Q^{\dag}\Psi=0$ in $\Omega$. Approach to a point on $X^{\hat{A}}_{\hat{i}}=(\mathrm{ctt})$. We have
\[\frac{\partial}{\partial X^A_i}\frac{\partial}{\partial\lambda^A_i}\Psi=0.
\]
Hence the relevant restriction for $\Psi_1$ is
\begin{equation}\frac{\partial}{\partial X^{\hat{A}}_{\hat{i}}}\frac{\partial}{\partial\lambda^{\hat{A}}_{\hat{i}}}\Psi=0.
\end{equation}
In explicit form we obtain
\begin{equation} \label{IV}  \frac{\partial \Psi^{\hat{i}}_{1\hat{A}}}{\partial X^{\hat{A}}_{\hat{i}}}=0.\end{equation}
From (\ref{II}) and (\ref{IV}), we get \[\frac{\partial\Psi^{j}_{1B}}{\partial X^{\hat{A}}_{\hat{i}}}=0\] for all $(B,j)$. Consequently
\[\frac{\partial\Psi^{j}_{1B}}{\partial X^{A}_{i}}=0
\]
for all $(B,j)$ and $(A,i)$ at the hypersurface $X^{\hat{A}}_{\hat{i}}=(\mathrm{ctt}).$ For the generic term in the expansion we get
\begin{equation}\frac{\partial\Psi^{\hat{i}j_2\dots j_m}_{m\hat{A}B_2\dots B_m}}{\partial X^{\hat{A}}_i}=0
\end{equation}
 which together with (\ref{III}) yields
\[\frac{\partial\Psi^{j_1j_2\dots j_m}_{B_1B_2\dots B_m}}{\partial X_i^A}=0\]
on the hypersurface defined by $X^{\hat{A}}_{\hat{i}}=(\mathrm{ctt})$ for any set of indices.

Therefore, according to the arguments in the previous two paragraphs, the condition $H\Psi=0$ in $\partial\Omega$, implies that \[\frac{\partial\Psi}{\partial X^A_i}=0 \qquad \mathrm{on}\quad \partial \Omega.\]

Now, the equation
\[H\Psi=0\] is an elliptic system of partial differential equations on the components of $\Psi$: $\Psi_0,\,\Psi_1,\,\Psi_2,\dots$.
On $\partial\Omega$ we have
\[\Psi_0=\Psi_1=\Psi_2=\dots=0\]
\[\partial_n\Psi_0=\partial_n\Psi_1=\partial_n\Psi_2=\dots=0.
\]
The partial differential system has analytic coefficients. Then, by virtue of the Cauchy-Kowalevski Theorem, it follows that $\Psi=0$ on $\Omega$.

A detailed argument along these lines can be rigorously established and will be reported in due course.

\section{Aknowledgements}  MPGM Would like to thank to the Theoretical Physics Department at U. Zaragoza, Spain, for kind invitation while part of this work was done. MPGM is supported by Mecesup ANT1398, Universidad de Antofagasta, (Chile). A.R. is partially supported by Projects Fondecyt
1121103 (Chile).


\section*{References}
\begin{thebibliography}{9}
\bibitem{dwhn}
B. de Wit, J. Hoppe, H. Nicolai, Nucl. Phys. {\bf B305} (1988) 545.
%
\bibitem{hoppe} J. Hoppe {\em On The Construction of Zero Energy States in Supersymmetric Matrix Models III} arXiv:hep-th/9711033
%
\bibitem{fh} J. Froehlich, J. Hoppe {\it On Zero-Mass Ground States in Super-Membrane Matrix Models} arXiv:hep-th/9701119
%
\bibitem{bfss}T. Banks, W. Fischler, S.H. Shenker, L. Susskind,
 Phys.Rev. D  (1997) {\bf 55}, 5112-5128
%
%
%
\bibitem{halpern} M.B. Halpern, C. Schwartz 1998 {\it  Int.J.Mod.Phys.} A{\bf 13} 4367
%
\bibitem{yi} P. Yi, Nucl. Phys. B 505,
307 (1997)
%
\bibitem{sethi-stern} S. Sethi and M. Stern,  Commun. Math. Phys. 194,
675 (1998)
%
\bibitem{porrati} M. Porrati, A. Rozenberg, Nucl. Phys. B515, 184-202 (1998)
%
\bibitem{hlt2} J. Hoppe, D. Lundholm, M. Trzetrzelewski
Nucl.Phys. {\bf B817} (2009) 155-166
%
\bibitem{maldacena} J M. Maldacena, Int.J.Theor.Phys. (1999){\bf 38}  1113-1133, Adv.Theor.Math.Phys.  (1998) {\bf 2},231-252
%
\bibitem{ydri} B. Ydri,
  Int.J.Mod.Phys.(2012) A{\bf 27}  1250088	
%
\bibitem{connor} D. O'Connor,
Theor.Math.Phys.(2011) {\bf 169}  1405-1412
%
\bibitem{dwln}
B. de Wit, M. Luscher, H. Nicolai, Nucl. Phys. {\bf B320} (1989) 135.
%
\bibitem{lin-yin}  Ying-Hsuan Lin, Xi Yin {\it  On the Ground State Wave Function of Matrix Theory} arXiv:1402.0055
%
\bibitem{fghhy} J. Frohlich, G.M. Graf, D. Hasler, J. Hoppe, Shing-Tung Yau, Nucl.Phys. {\bf B567} (2000) 231-248
%
\bibitem{su(2)} A.M. Khvedelidze, H.P. Pavel
Phys.Lett.A  (2000) {\bf 267} 96-100
%
\bibitem{hp}J. Hoppe, J. Plefka {it The Asymptotic groundstate of SU(3) matrix theory}
 hep-th/0002107
%
\bibitem{dd}J. Daboul, R. Delbourgo,
 J.Math.Phys. (1999){\bf 40}  4134-4150
%
\bibitem{oqm}S. De Leo, Khaled Abdel-Khalek,
Prog.Theor.Phys.(1996) {\bf 96}  823-832
\bibitem{jordan}P. Jordan, Z. Phys.
80, 285 (1933); P. Jordan and E. P. Wigner, Ann. Math.
35, 29 (1934)
%
\bibitem{veiro-restuccia} A. Restuccia and  J.P. Veiro,
{\it On the Formulation of Yang-Mills Theory with the Gauge Field Valued on the Octonionic Algebra}
arXiv:1412.4889
%
\bibitem{hlt} J. Hoppe, D. Lundholm, M. Trzetrzelewski {\it Annales Henri Poincare} 2009 {\bf 10} 339
%
\end{thebibliography}
\end{document}